\documentclass[12pt]{iopart}
\usepackage{times}
\usepackage{cite}
\usepackage{epsf}

\begin{document}
\begin{flushright}
MC-TH-2002-08\\
hep-ph/0210266\\
October 2002
\end{flushright}

\title{Minimal Nonminimal Supersymmetric Standard Model}
\author{C. Panagiotakopoulos$^a$ and A. Pilaftsis$^b$\footnote[1]{Talk
given at the conference ``Beyond the Desert 2002,'' 2--7 June 2002,
Oulu, Finland} }
\medskip
\address{
$^a$  Physics Division, School of Technology, 
      Aristotle University of Thessaloniki,\\
      54006 Thessaloniki, Greece\\[3mm]
$^b$  Department of Physics and Astronomy, 
      University of Manchester,\\ 
      Manchester M13 9PL, United Kingdom    }
\begin{abstract}
We review the basic  field-theoretic and phenomenological  features of
the recently  introduced  Minimal Nonminimal  Supersymmetric  Standard
Model~(MNSSM). The introduced model is the  simplest and most economic
version among the  proposed nonminimal supersymmetric models, in which
the so-called $\mu$-problem  can be successfully addressed. As opposed
to  the MSSM and   the    frequently-discussed NMSSM, the  MNSSM   can
naturally predict the existence of a light  charged Higgs boson with a
mass smaller  than 100~GeV. Such a possible  realization of  the Higgs
sector  can  be soon  be tested  at the upgraded  Run  II phase of the
Tevatron collider.
\end{abstract}

It is known that  Minimal Supersymmetric Standard Model~(MSSM) suffers
from  the so  called $\mu$-problem.  The   superpotential of  the MSSM
contains a  bilinear   term $-\mu      \widehat{H}_{1}\widehat{H}_{2}$
involving   the two  Higgs-doublet superfields  $\widehat{H}_{1}$  and
$\widehat{H}_{2}$,  known as the  $\mu$-term.  Naive implementation of
the $\mu$-parameter within supergravity theories would lead to a $\mu$
value of the order of the Planck scale  $M_{{\rm P}}$.  However, for a
successful    Higgs mechanism at      the  electroweak   scale,    the
$\mu$-parameter is actually required  to  be many orders of  magnitude
smaller  of   order $M_{{\rm SUSY}}$.   Many  scenarios,  all based on
extensions    of the  MSSM,   have  been  proposed   in the   existing
literature~\cite{mu} to provide  a natural explanation for  the origin
of the $\mu$-term.

Recently,   a       minimal   extension of     the      MSSM  has been
presented~\cite{PT,PP1,DHMT,PP2}, called the      Minimal   Nonminimal
Supersymmetric Standard  Model~(MNSSM)~\cite{PP1,PP2}, in which    the
$\mu$-problem can be successfully  addressed in a rather  minimal way.
In the MNSSM the   $\mu$-parameter is promoted   to a  chiral  singlet
superfield $\widehat{S}$,  and   all  linear,  quadratic   and   cubic
operators  involving    only   $\widehat{S}$ are    absent  from   the
renormalizable superpotential; $\widehat{S}$ enters through the single
term $\lambda\, \widehat{S}\,\widehat{H}_1 \widehat{H}_2$:
\begin{equation}
  \label{Wren}
W_{\rm MNSSM}^{\rm ren}\ =\ \widetilde{W}_{\rm MSSM}\: 
+\: \lambda\, \widehat{S}\,
\widehat{H}^T_1\, i\tau_2\,\widehat{H}_2\, ,
\end{equation}
where  $\widetilde{W}_{\rm MSSM}$  is the  superpotential of  the MSSM
without  the  presence of  the  $\mu$  term.   The crucial  difference
between  the   MNSSM  and  the   frequently-discussed  Next-to-Minimal
Supersymmetric  Standard Model~(NMSSM)~\cite{nmssm}  lies in  the fact
that  the  cubic  term  $\frac{1}{3}\kappa\, \widehat{S}^3$  does  not
appear in the renormalizable superpotential of the former.

The  key  point in  the   construction   of the renormalizable   MNSSM
superpotential is that the simple form~(\ref{Wren}) may be enforced by
discrete $R$-symmetries, such as ${\cal Z}^R_5$~\cite{PT,PP1,DHMT,PP2}
and  ${\cal Z}^R_7$~\cite{PP1,PP2}.     These discrete $R$-symmetries,
however,  must be extended  to the  gravity-induced non-renormalizable
superpotential and  K\"ahler  potential terms as well.  To communicate
the breaking of  supersymmetry  (SUSY), we  consider  the  scenario of
$N=1$ supergravity spontaneously    broken by a set of   hidden-sector
fields at   an    intermediate scale.    Within   this   framework  of
SUSY-breaking, we have then  been   able to show~\cite{PP1} that   the
above $R$-symmetries are sufficient to postpone  the appearance of the
potentially  dangerous tadpole~\cite{NIL,Bag}   ~$t_S\, S$  at  a loop
level $n$ higher than 5, where
\begin{equation}
  \label{tS}
t_S\ \sim\ \frac{1}{(16\pi^2)^n}\ M_{\rm P}\,M^2_{\rm SUSY}\; .
\end{equation}
{}From this last  expression, one  can  estimate that the  size of the
tadpole parameter $t_S$    is  in  the  right  ballpark,   i.e.~$|t_S|
\stackrel{<}{{}_\sim} 1$--10~TeV$^3$ for  $n  =  6,7$, such that   the
gauge hierarchy does not get destabilized. To be specific, the tadpole
$t_S\, S$ together with the soft SUSY-breaking mass  term $m^2_S S^* S
\sim M^2_{\rm  SUSY} S^* S$ lead  to a vacuum  expectation value (VEV)
for $S$,  $\big<  S\big> = \frac{1}{\sqrt{2}}   v_S$, of order $M_{\rm
SUSY} \sim 1$~TeV.  The latter gives  rise to a $\mu$-parameter at the
required electroweak scale, i.e.\
\begin{equation}
\mu\ =\ -\frac{1}{\sqrt{2}}\,    \lambda v_S\ \sim\  M_{\rm SUSY}\; .
\end{equation}  
Thus, a natural explanation for the  origin of the $\mu$-parameter can
be obtained.  Finally, since  the   effective tadpole term  $t_S\,  S$
explicitly breaks the  continuous Peccei--Quinn symmetry governing the
remaining renormalizable Lagrangian of the MNSSM, the theory naturally
avoids the presence of a phenomenologically excluded weak-scale axion.

\begin{figure}[t]
   \leavevmode
\begin{flushleft}
   \epsfxsize=17cm
   \epsfysize=19.5cm
    \epsffile{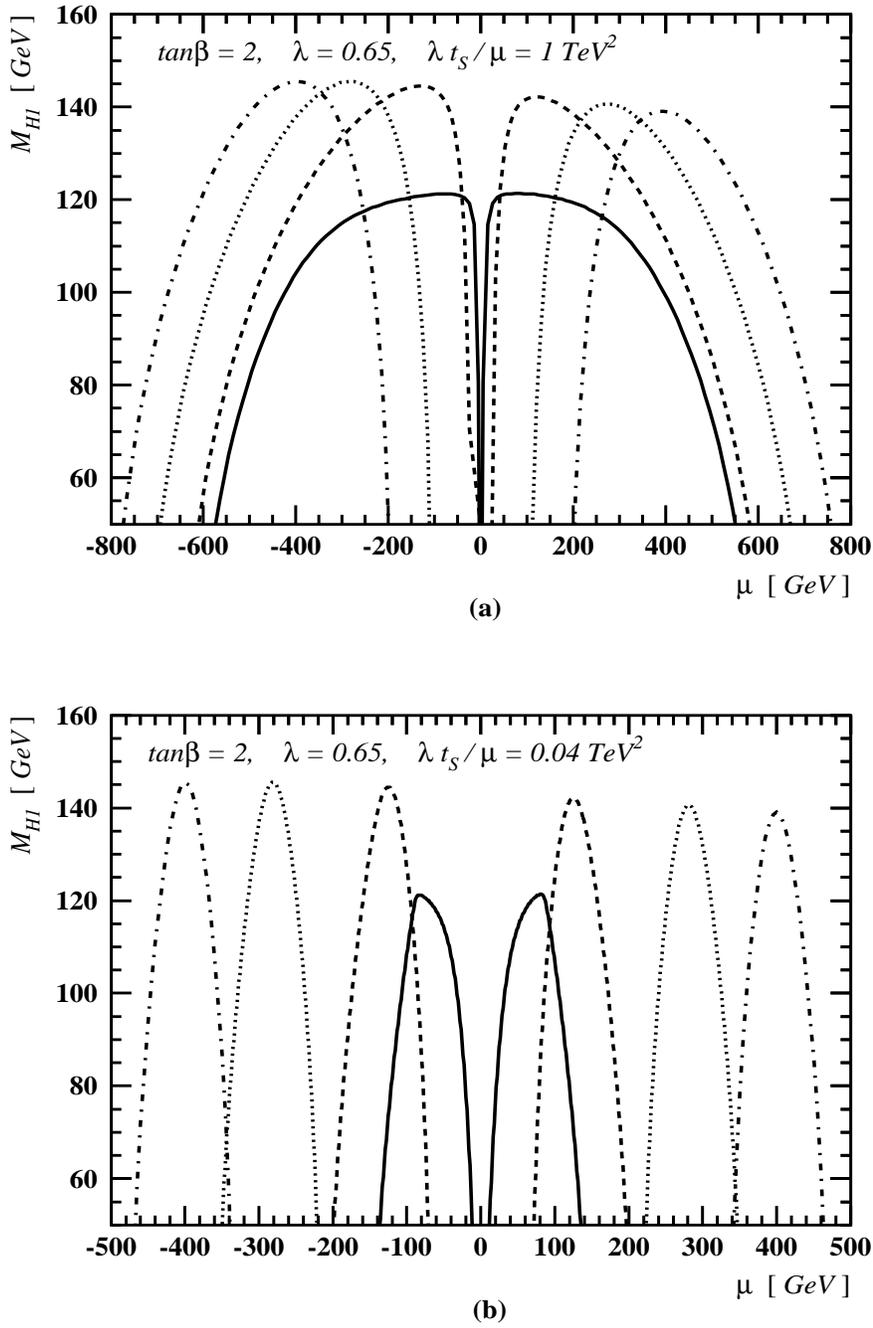}
\end{flushleft}
\vspace{-1.5cm}
\caption{Numerical values for $M_{H_1}$ versus $\mu$ in the MNSSM with
$m^2_{12} = 0$, for $M_{H^+} = 0.1$~(solid), 0.3 (dashed),
0.7~(dotted) and 1~(dash-dotted)~TeV.}\label{fig1}
\end{figure}

\begin{figure}[t]
   \leavevmode
 \begin{center}
   \epsfxsize=16cm
    \epsffile[0 50 567 454]{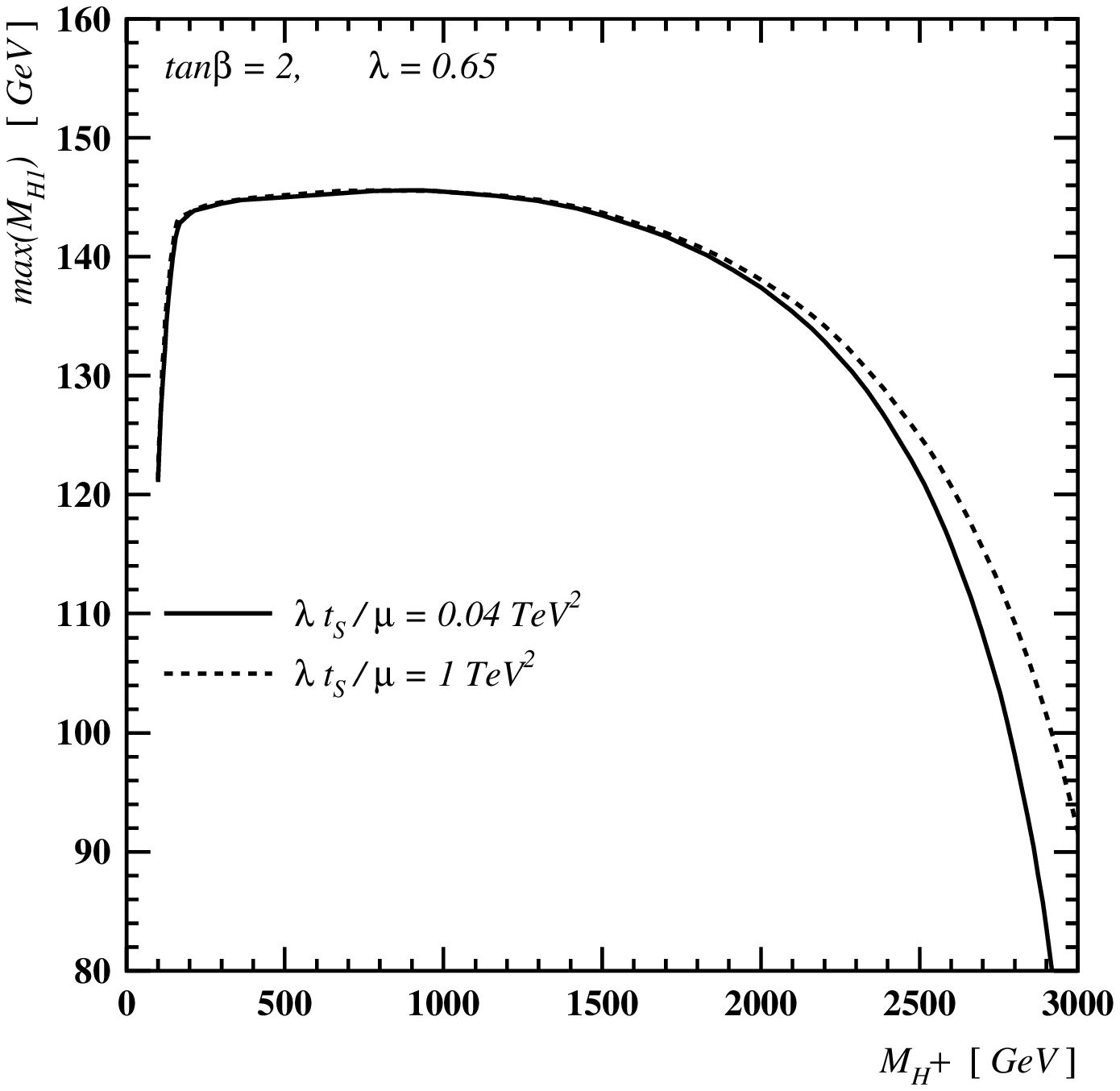}
 \end{center} 
\caption{The maximal predicted value of $M_{H_1}$ 
  as a function of the charged Higgs-boson mass $M_{H^+}$ in the MNSSM
  with $m^2_{12} =0$.}\label{fig2}
\end{figure}

In addition   to  the tadpole $t_S$ of   the  physical scalar  $S$, an
effective      tadpole   for   its   auxiliary    component   $F_S$ is
generated~\cite{Bag}.  However, depending  on the underlying mechanism
of SUSY breaking, the effective tadpole proportional to $F_S$ could in
principle  be absent   from  the model.    Such   a reduction  of  the
renormalizable operators does not thwart  the renormalizability of the
theory.  The   resulting renormalizable low-energy   scenario has  one
parameter  less than the    frequently-discussed NMSSM with  the cubic
singlet-superfield  term $\frac{\kappa}{3}\widehat{S}^3$  present;  it
therefore represents  the most economic, renormalizable  version among
the non-minimal supersymmetric models proposed in the literature.
  
As  opposed to the NMSSM, the  MNSSM satisfies the tree-level mass sum
rule~\cite{PP1}:
\begin{equation}
  \label{sumrule}
M^2_{H_1}\: +\: M^2_{H_2}\: +\: M^2_{H_3}\ =\ M^2_Z\: +\: M^2_{A_1}\:
+\: M^2_{A_2}\,,
\end{equation}
where $H_{1,2,3}$ and  $A_{1,2}$ are the  three CP-even and two CP-odd
Higgs     fields,  respectively.      The     tree-level   mass    sum
rule~(\ref{sumrule}) is very analogous to the corresponding one of the
MSSM \cite{GT}, where the two heavier Higgs states $H_3$ and $A_2$ are
absent in the latter.  This striking analogy  to the MSSM allows us to
advocate that the  Higgs sector of the  MNSSM differs indeed minimally
from the one of the MSSM, i.e.\ the introduced model truly constitutes
the minimal  supersymmetric extension of  the MSSM.  In the NMSSM, the
violation of the mass sum  rule~(\ref{sumrule}) can become much larger
than the one induced by  the one-loop stop/top effects, especially for
relatively large values of $|\kappa |$, $|\mu|$ and $|A_\kappa|$.
  
In the non-minimal supersymmetric  standard models, the upper bound on
the lightest  CP-even Higgs-boson   mass $M_{H_1}$ has   a  tree-level
dependence on the coupling $\lambda$~\cite{nmssm,EQ,PP1,DHMT}, i.e.
\begin{equation}
  \label{mbound}
M^{2(0)}_{H_1} \ \le\ M^2_Z\, \bigg( \cos^2 2\beta\:
+\: \frac{2\,\lambda^2}{g^2_w + g'^2}\, \sin^2 2\beta\, \bigg)\,,
\end{equation}
where the angle $\beta$ is  defined by means of $\tan\beta = v_2/v_1$,
the ratio of  the VEVs of the two Higgs doublets.   Since in the MNSSM
$\lambda$ can  take its maximum allowed  value naturally corresponding
to the  NMSSM with $\kappa =  0$~\cite{EQ}, the value  of $M_{H_1}$ is
predicted     to    be    the     highest.     In     particular,    a
renormalization-group-improved  analysis~\cite{PP2}  of the  effective
MNSSM   Higgs   potential  leads   to   the   upper  bound:   $M_{H_1}
\stackrel{<}{{}_\sim}  145$~GeV,  for  large  stop  mixing  (see  also
Fig.~\ref{fig1}). Consequently, such a scenario can only be decisively
tested  by the  upgraded  Run II  phase  of the  Tevatron collider  at
Fermilab and by the Large Hadron Collider (LHC) at CERN.

\begin{figure}[t]
   \leavevmode
 \begin{center}
   \epsfxsize=17cm
   \epsfysize=19.5cm
    \epsffile{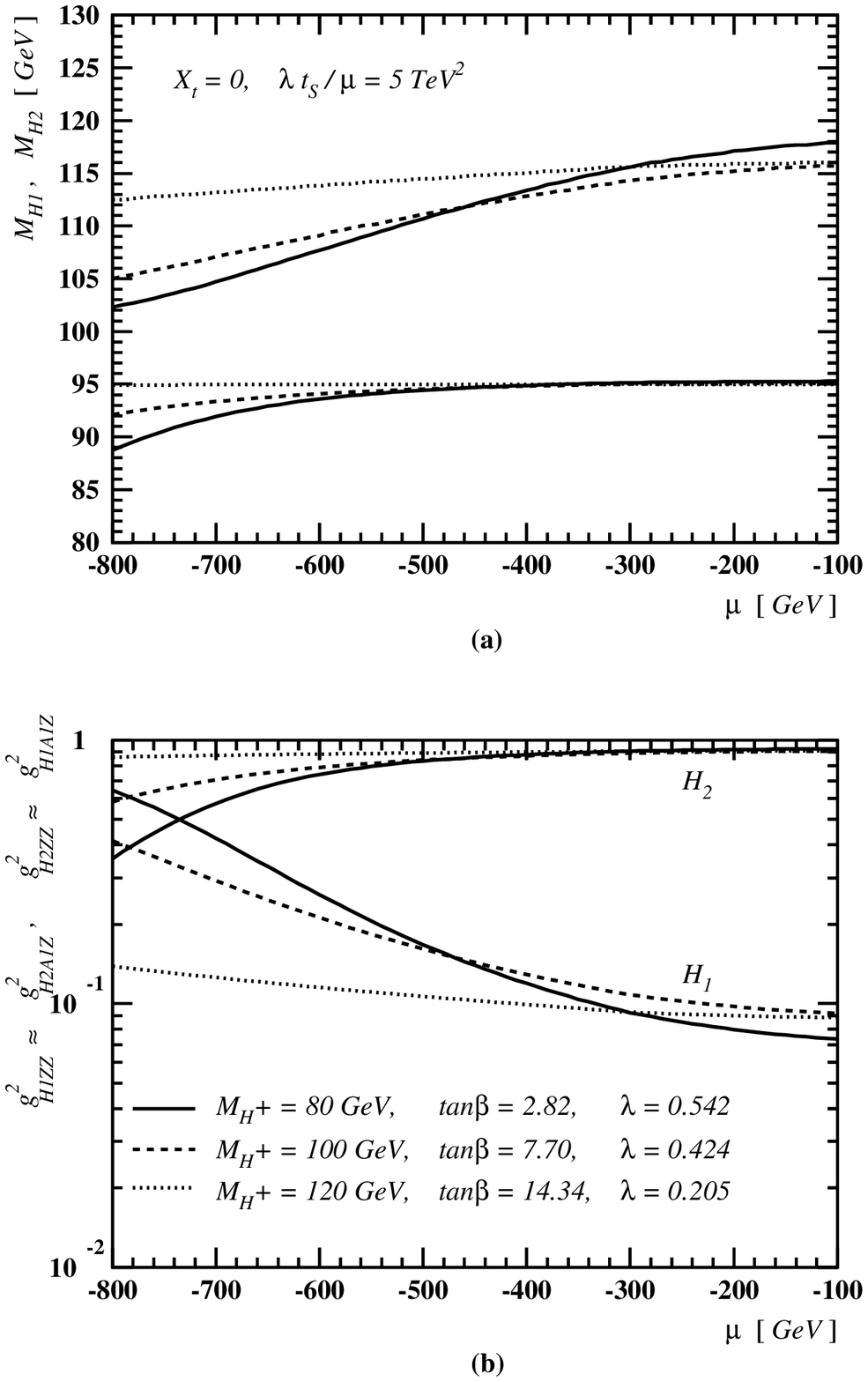}
 \end{center}
\vspace{-1.5cm} 
\caption{Numerical predictions for (a) $M_{H_1}$ and $M_{H_2}$,
  and (b) $g^2_{H_1ZZ}$ and  $g^2_{H_2ZZ}$,  as functions of $\mu$  in
  the MNSSM.}\label{fig3}
\end{figure}

\begin{figure}[t]
   \leavevmode
 \begin{center}
   \epsfxsize=17cm
   \epsfysize=19.5cm 
    \epsffile{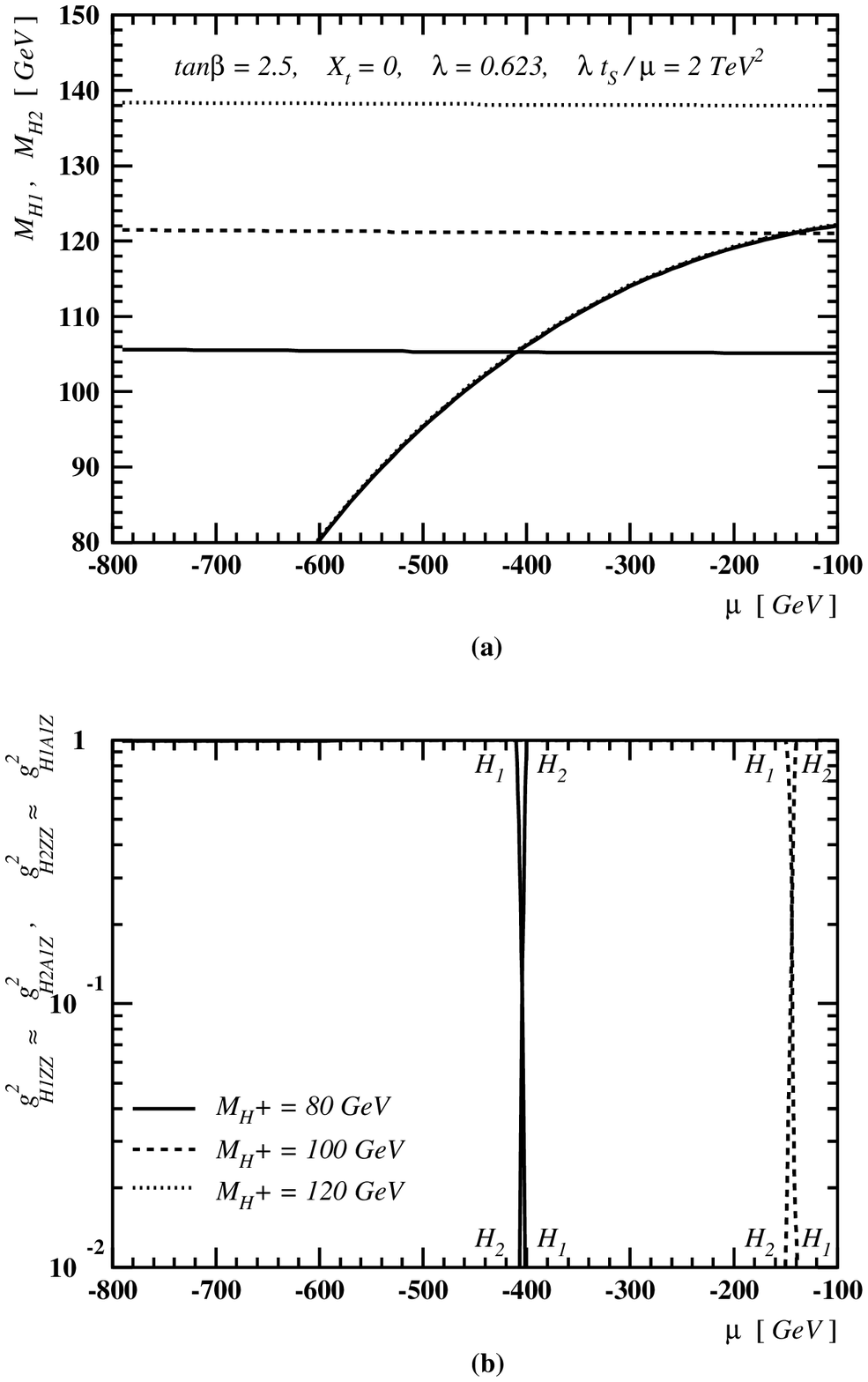}
 \end{center}
\vspace{-1.5cm} 
\caption{Numerical values of (a) $M_{H_1}$ and $M_{H_2}$,
  and (b) $g^2_{H_1ZZ}$ and  $g^2_{H_2ZZ}$,  as functions of $\mu$  in
  the MNSSM.}\label{fig4}
\end{figure}
 
The MNSSM can comfortably predict  viable scenarios, where the mass of
the charged Higgs   boson $H^+$  is in  the  range: 80~GeV~$<  M_{H^+}
\stackrel{<}{{}_\sim}$~3~TeV, for phenomenologically  relevant  values
of $|\mu | \stackrel{>}{{}_\sim} 100$ GeV~\cite{GKLR}. In fact, as can
be seen from~Fig.~\ref{fig2}, there   is  an absolute upper bound   on
$M_{H^+}$ for fixed values of $\lambda$ and $\tan\beta$. 

On the other hand, charged Higgs-boson masses smaller than 100~GeV can
naturally be obtained within the  MNSSM, while the SM-like Higgs boson
$H_{\rm SM}$, with dominant coupling to the $Z$  boson, can be heavier
than about 115~GeV~\cite{PP1,PP2}.  Instead, both  in the MSSM and the
NMSSM~\cite{PP1}, such a  Higgs-boson mass  spectrum is  theoretically
inaccessible,  if  the  phenomenologically favoured   range  $|\mu   |
\stackrel{>}{{}_\sim} 100$ GeV is considered.  In Figs.~\ref{fig3} and
\ref{fig4}, we display numerical values for the masses of the lightest
and   next-to-lightest Higgs bosons,     $H_1$  and $H_2$,  and  their
couplings  to  the $Z$ boson as  functions  of $\mu$,  for a number of
versions of the MNSSM that predict light charged Higgs bosons.  In the
MNSSM versions  under  study, the  SM-like  Higgs  boson  $H_{\rm SM}$
(mainly  $H_2$) can have a  mass  larger 110~GeV,  compatible with the
present experimental bound.  The generic  prediction is that the first
CP-even  Higgs boson $H_1$ is lighter  than $H_2$ and has a suppressed
coupling   to the $Z$  boson in    agreement with  LEP2  data.  {}From
Figs.~\ref{fig3} and~\ref{fig4},  we  observe  that the  charged Higgs
boson   can be   as light  as  the  present  experimental upper bound,
i.e.~$M_{H^+}\sim  80$~GeV.  This  is   an important  phenomenological
feature of the MNSSM,  which is very helpful  to discriminate it  from
the NMSSM. It  is a reflection of  a new non-trivial decoupling  limit
due to a  large tadpole   $|t_S|$,  which is  only attainable  in  the
MNSSM~\cite{PP1}.  In this limit, the heavier Higgs states $H_{3}$ and
$A_3$ can both  decouple from the Higgs spectrum  as a heavy singlets.
The  upcoming upgraded Run~II phase of  the  Tevatron collider has the
physics      potential  to      probe     the     viability    of    a
light-charged-Higgs-boson realization.

For  scenarios with   $M_{H^+}  \stackrel{>}{{}_\sim}  200$  GeV,  the
distinction  between the MNSSM  and  the NMSSM becomes more difficult.
In this case,  additional experimental information  would be necessary
to distinguish  the two SUSY extensions  of the MSSM, resulting from a
precise determination of the masses,  the widths, the branching ratios
and   the production cross sections  of   the CP-even and CP-odd Higgs
bosons.   Nevertheless, if  the  tadpole  parameter $\lambda  t_S/\mu$
becomes much larger than $M^2_{H^+}$    with the remaining   kinematic
parameters held  fixed, the  Higgs   states $H_3$ and $A_2$   will  be
predominantly singlets.   As an important phenomenological consequence
of this, the  complementarity  relations between the $H_{1,2}ZZ$-  and
$H_{2,1}A_1Z$-  couplings will  then  hold approximately   true in the
MNSSM, i.e.
\begin{equation}
  \label{compl}
g^2_{H_1ZZ}\ =\ g^2_{H_2A_1 Z}\,,\qquad g^2_{H_2 ZZ}\ =\ g^2_{H_1 A_1 Z}\; .
\end{equation}
In addition, the couplings of the  two heaviest states $H_3$ and $A_2$
to  the gauge  bosons  will vanish.  Here,  we  should stress that the
relations~(\ref{compl}) are not generically   valid in the  NMSSM. The
latter is  a consequence of   the absence of the aforementioned  large
tadpole decoupling limit, such that  the states $H_3$ and $A_2$  could
decouple as singlets. Future next linear $e^+  e^-$ colliders have the
capabilities   to   experimentally   determine  the  $H_{1,2}ZZ$-  and
$H_{2,1}A_1Z$- couplings to an accuracy even up to 3\% and so test, to
a  high degree, the  complementarity relations~(\ref{compl}) which are
an essential phenomenological feature of the MNSSM.

As  has been  discussed  in~\cite{PP1}, the  MNSSM also  predicts  the
existence     of a   light  neutralino, the    axino.    The axino  is
predominantly a singlet  field, for $|\mu| \stackrel{>}{{}_\sim}  120$
GeV.  LEP  limits    on the  $Z$-boson invisible   width   lead to the
additional     constraint:    $200   \stackrel{<}{{}_\sim}    |\mu   |
\stackrel{<}{{}_\sim} 250$ GeV,  for $\lambda \approx 0.65$.  However,
such  a  constraint disappears  completely     for smaller values   of
$\lambda$, namely for  $\lambda \stackrel{<}{{}_\sim} 0.45$.  In fact,
the axino may become   the lightest supersymmetric particle,  which is
very long-lived in the MNSSM, and hence it  potentially qualifies as a
candidate for cold dark matter. We feel that a dedicated study in this
direction needs to be done.

Let us   summarize  the  basic  field-theoretic   and phenomenological
features of the  MNSSM: (i) The MNSSM minimally  departs from the MSSM
through  the  presence   of  a  gauge-singlet  superfield whose    all
self-couplings  are absent.  On  the basis of discrete $R$ symmetries,
such as ${\cal Z}^R_5$ and ${\cal Z}^R_7$, the quadratically divergent
harmful tadpoles  first appear  at the 6-  and  7-loop levels, thereby
avoiding to  destabilize the gauge  hierarchy.  By the same token, the
MNSSM  can minimally account for  the origin of $\mu$-term; (ii) Since
the loop-induced tadpoles break any  continuous or discrete  symmetry,
the model does not suffer  from problems~\cite{ASW} related to visible
weak-scale axions  and domain walls; (iii)  As a consequence of  a new
decoupling  limit due  to a  large   tadpole, the MNSSM can  naturally
predict viable scenarios in  which the  charged  Higgs boson  $H^+$ is
much lighter  than the neutral Higgs  boson with a SM-type coupling to
the $Z$  boson.  The  planned  colliders, i.e.\ the  upgraded Tevatron
collider~\cite{revII} and the LHC, have  the potential capabilities to
test such interesting  scenarios with a  relatively  light $H^+$; (iv)
Unlike the  frequently-discussed NMSSM, the  Higgs sector of the MNSSM
exhibits a much closer resemblance to the one of the MSSM, by means of
the tree-level mass sum  rule~(\ref{sumrule}) and the  complementarity
relations~(\ref{compl}) of the Higgs-boson couplings to the $Z$ boson.

In conclusion, all the above  facts point to a single perspective: the
only truly minimal supersymmetric extension of the MSSM is the Minimal
Nonminimal Supersymmetric Standard Model.

\end{document}